# Assessment of Performance of Correlation Estimates in Discrete Bivariate Distributions using Bootstrap Methodology

**Michael Tsagris[1], Ioannis Elmatzoglou[2], Christos C. Frangos[3]**

[1]School of Mathematical Sciences, University of Nottingham, UK

[2]Zeincro S.A., Athens, Greece

[3]Department of Business Administration, Technological Educational Institute of Athens, Greece

Corresponding author:

Michael Tsagris

School of Mathematical Sciences, Division of Statistics

University of Nottingham

University Park, Nottingham NG7 2RD

United Kingdom

Tel.: +44 (0) 7503242099, Fax: +44 (0) 115 95 13837

e-mail: pmxmt1@nottingham.ac.uk

**ABSTRACT**

Little attention has been given to the correlation coefficient when data come from discrete or continuous non-normal populations. In this paper we considered the efficiency of two correlation coefficients which are from the same family, Pearson's and Spearman's estimators. Two discrete bivariate distributions were examined, the Poisson and the Negative Binomial. The comparison between these two estimators took place using classical and bootstrap techniques for the construction of confidence intervals. Thus, these techniques are also subject to comparison. Simulation studies were also used for the relative efficiency and bias of the two estimators. Pearson's estimator performed slightly better than Spearman's.

**1. INTRODUCTION**

Bootstrap and jackknife are well-known resampling methodologies for obtaining nonparametric confidence intervals of a parameter. In most statistical problems one needs an estimator of an unknown parameter of interest as well as some assessment of its variability. In many such problems the estimators are complicated functionals of the empirical distribution function and it is difficult to derive trustworthy analytical variance estimates for them. Bootstrap (Efron, 1979, 1982, 1987) and jackknife (Miller, 1974) use straightforward but extensive computing to produce reliable indications of the variability of an estimator. For the justification of bootstrap with regard to the foundation of its theoretical basis we have to say the following. The primary objective of this technique is to estimate the sampling distribution of a statistic. Essentially, bootstrap is a method that mimics the process of sampling from a population, like one does in Monte Carlo simulations, but instead drawing samples from the observed sampling data. The tool of this mimic process is the Monte Carlo algorithm of Efron (1979).

This process is explained properly by Efron and Tibshirani (1993), who also noted that bootstrap confidence intervals are approximate yet better than the standard ones. Nevertheless, they do not try to replace the theoretical ones and neither is bootstrap a substitute for precise parametric results, but rather a way to reasonably proceed when such results are unavailable.

Bootstrap is in line with the general direction which Rao has imposed for the statistical discipline and is "to extract all the possible information from the data" (Rao, 1989). One, however, can pose the following question: Why is the bootstrap estimate of variance converging to the true variance? The answer is the following: The strong law of large numbers implies that $\hat{S}_B^2 \to S^2$ almost surely as $B \to \infty$, in which case B stands for the number of bootstrap samples and $S$ stands for the standard deviation. What is the permissible necessary value of B? In this respect we use the reference of Booth and Sarkar (1998). These researchers suggested that a choice between 200 and 800 is satisfactory for the estimation of the standard error and thus for the construction of confidence intervals. With the great power of computing today we can have a safe choice of B equal to 1000, although Chernick (2008) suggests using B=100,000.

In this paper we have not used powerful bootstrap variations like the partial likelihood approach and bootstrap calibration. These methods have been proposed by Davison et al. (1992), Beran (1987), Loh (1987) and Efron and Tibshirani (1993).

The choice for study of correlation coefficients was due to the fact that not much attention has been given to this parameter and most of the authors who have dealt with it used bivariate normal or log-normal distributions. Efron and Tibshirani (1993) used bootstrap techniques to estimate the standard error of (Pearson's) rho in the bivariate normal distribution. They also showed that the non-parametric delta method can be badly biased. Dolker et al. (1982) pointed out some possible problems in the correlation coefficient for very small sample sizes. The probability of such discrepancies (same bivariate vectors) is very low in our case.

Rasmussen (1987) worked on the estimation of Pearson's correlation coefficient for normal and non-normal distributions and saw that Pearson's coefficient is robust to deviations from the normality assumption. Hall et al. (1989), Frangos and Schucany (1990) and Lee and Rodgers (1998) studied the bivariate normal and log-normal distributions. In all cases, continuous bivariate distributions, and mainly the normal distribution, were examined. There are more examples of researchers who studied the correlation coefficient. For instance, Young (1988) used the Gaussian, log-normal and t distributions, and Hall et al. (1989) also used normal and log-normal distributions but with small sample sizes (n=8, 10 and 12). Lunneborg (1985) used a sample of the famous LSAT data for which the distribution is close to bivariate normal.

In this paper the correlation coefficient is examined in the bivariate Poisson and Negative Binomial distributions. The disadvantage of these two bivariate distributions is that the correlation coefficient is strictly non negative. However, the bivariate Poisson has a nice property. When the covariance term is zero and hence the correlation coefficient is zero, we end up with two independent Poisson variates (Kawamura, 1973). Both of these discrete distributions have a wide variety of applications, such as sports (Karlis and Ntzoufras, 2003, 2005) or medicine (hospital and doctor visits for instance). For more information about the applications of the Poisson or Negative Binomial models one can also read the two papers by Greene (2007).

Section 2 of this paper describes the Normal, Basic, Percentile, ABC, Studentized and the accelerated bootstrap (BCa) methods, using both the positive and negative jackknives to estimate the acceleration constant a. Section 3 describes the Classical method of constructing confidence intervals for the correlation coefficient. Section 4 describes some characteristics of these confidence intervals. The comparison is accomplished by an extensive simulation study in which the properties of these intervals are examined in sections 4 and 5. All algorithms were implemented in R 2.9.0 and the package "boot" was used.

**2. NON PARAMETRIC BOOTSTRAP CONFIDENCE INTERVALS**

Let $X_1, X_2,...,X_n$, be n independent and identically distributed random variables from an unknown probability distribution $F_\theta(x)$. Let $\hat{F}$ be the empirical distribution function, having mass $1/n$ at each observed $x_i$. The essence of bootstrap methodology is that one draws random samples $X_1^*, X_2^*,..., X_n^*$ with replacement from $\hat{F}$ and then calculates $\hat{\theta}^* = \hat{\theta}(X_1^*,...,X_n^*)$ as an estimate of θ. After B replications of this process, one has an empirical distribution of $\hat{\theta}_b^*$ (b=1,…,B) values, which serves as an estimate of the unknown sampling distribution of $\hat{\theta}$.

Following Efron (1987), to construct nonparametric confidence intervals for θ via the Percentile method (PM), one uses the 100α and 100(1-α) percentiles of the bootstrap distribution of θ. The 100(1-2α)% central confidence interval for θ is given by

$$\theta \in [\hat{G}_B^{-1}(a), \hat{G}_B^{-1}(1-a)] \tag{2.1}$$

with
$$\hat{G}_B(t) = \frac{\#\{\hat{\theta}_b^* \leq t\}}{B}$$

(2.2)

being the estimated bootstrap distribution function.

Normal method (or standard confidence interval) has the known form of

$$\theta \in [\hat{\theta} - Z^{(1-\alpha)} \cdot SE(\hat{\theta}), \hat{\theta} + Z^{(\alpha)} \cdot SE(\hat{\theta})] \tag{2.3}$$

where $SE(\hat{\theta})$ is a reasonable estimate of the standard error of $\hat{\theta}$ based upon the $\hat{\theta}_b^*$ (b=1,…,B) values and $Z^{(\alpha)} = \Phi^{-1}(\alpha)$ can be obtained from the tables of standard normal distribution.

The bias-corrected method with acceleration constant α, (BCα), introduced by Efron and Tibshirani (1986) and discussed in detail by Efron (1987), is a procedure for improved confidence intervals for problems where there exists a monotone transformation g such that $\hat{\phi} = g(\hat{\theta})$ and $\phi = g(\theta)$ satisfy the approximation $\hat{\phi} \sim N(\phi - Z_0 \sigma_\phi, \sigma_\phi^2)$, where $\sigma_\phi = 1 + \alpha\phi$.

This yields the 100(1 – 2α)% confidence interval

$$\theta \in \{G_B^{-1}(\Phi(Z[\alpha])), G_B^{-1}(\Phi(Z[1-\alpha]))\} \tag{2.4}$$

where

$$Z[\alpha] = \hat{Z}_0 + \frac{\hat{Z}_0 - Z_\alpha}{1 - \hat{\alpha}(\hat{Z}_0 - Z_\alpha)} \tag{2.5}$$

Note that Bias Correction (BC) intervals are BCα, with $\hat{\alpha} = 0$ and they further reduce to PM when $\hat{Z}_0 = 0$. The bias correction $\hat{Z}_0$ is calculated using the formula $\hat{Z}_0 = \Phi^{-1}\left(\frac{\#\{\hat{\theta}_b^* < \hat{\theta}\}}{B}\right)$.

How does one find $\hat{\alpha}$? Efron (1987) showed that for one-parameter families, $f_\theta(T)$, of sampling distributions of

$T = \hat{\theta}$, a good approximation is

$$\hat{\alpha} = \frac{1}{6} \text{SKEW}_{\theta=\hat{\theta}}(i_\theta(T)) \tag{2.6a}$$

where $SKEW_{\theta=\hat{\theta}}(i_\theta(T))$, is the skewness of Fisher's score function $i_\theta(T) = (\partial/\partial\theta)\ln f_\theta(T)$ at $\theta = \hat{\theta}$. He also proposed that for data from an arbitrary distribution $F_\theta$ and $\theta = t(F_\theta)$ the constant $\hat{\alpha}$ is reasonably well approximated by

$$\hat{\alpha} = \frac{1}{6} \cdot \frac{\sum_{i=1}^{n}(I_i)^3}{\left\{\sum_{i=1}^{n}(I_i)^2\right\}^{3/2}} \tag{2.6b}$$

where $I_i$ is the influence function of the functional t at the point $x_i$. Two different finite sample estimates of the influence function $I_i$ are investigated here: the negative jackknife ($I_-$):

$$I_{i-} = (n-1)[(\hat{\theta}(x_1,...,x_n) - \hat{\theta}(x_1,...,x_{i-1},x_{i+1},...,x_n)] \tag{2.7a}$$

and the positive jackknife ($I_+$):

$$I_{i+} = (n+1)[(\hat{\theta}(x_1,...,x_n,x_i) - \hat{\theta}(x_1,...,x_n)]. \tag{2.7b}$$

The ABC method (stands for *approximate bootstrap confidence* intervals) is a method of approximating the BCα interval endpoints analytically, without using any Monte Carlo replications at all (Efron and Tibshirani, 1993). It requires the resampling vector $P^* = (P_1^*, P_2^*, ..., P_n^*)$ which consists of the proportions

$$P_i^* = N_i^*/n = \frac{\#\{x_j^* > x_i^*\}}{n}, \quad i=1,2,\ldots,n.$$ Therefore, the statistic e.g. $\hat{\theta}^* = \sum_{i=1}^{n} A_i^*$ is expressed as $\hat{\theta}^* = \sum_{i=1}^{n} P_i^* A_i$. The calculation of the confidence limits requires the calculation of some empirical influence components and of the acceleration constant which is again calculated as 1/6 times the standardized skewness of the empirical influence components.

Following Abramovitch and Singh (1985), as well as Loh (1987), we investigate a studentized statistic that is resampled to yield bootstrap confidence intervals for $\theta$. The approximation of the distribution of $(\hat{\theta} - \theta)/SE(\hat{\theta})$ was carried out using the bootstrap distribution of the studentized pivotal quantity (SPQ), namely

$$t^* = (\hat{\theta}^* - \hat{\theta})/SE(\hat{\theta}^*), \qquad (2.8)$$

where $SE(\hat{\theta})$ is the square root of the estimated $Var(\hat{\theta})$. Denote by $G_s(t) = \#\{t_b^* < t\}/B$ the bootstrap distribution of the studentized quantity $t^*$. The bootstrap confidence interval has the form:

$$\theta \in [\hat{\theta} - G_s^{-1}(1-a)SE(\hat{\theta}), \ \hat{\theta} - G_s^{-1}(a)SE(\hat{\theta})] \qquad (2.9)$$

The estimator for the standard error (the denominator in 2.8) of Pearson's correlation coefficient is $(1-\hat{\theta}^2)/\sqrt{n-3}$. The assumption of normality is necessary in order for this estimator to hold true, even though this assumption is not realistic at all. As for Spearman's correlation coefficient, a different approach was used. Instead of performing a second bootstrap in each bootstrap sample to estimate the standard error we estimated this standard error using simulation. That is, for every combination of sample size and value for the correlation coefficient we generated 1000 pairs, estimated the Spearman correlation coefficient and then calculated the standard deviation of these values.

The Basic method is a combination of the Percentile and Studentized methods. Instead of trying to find the empirical distribution of $\hat{\theta}_b^*$ (b=1,…,B), this method finds the empirical distribution of $\hat{\theta}_b^* - \hat{\theta}$. Note that this statistic is the same as in the studentized method with the only difference being that the denominator (standard error of $\hat{\theta}$) is set equal to one). The confidence interval for the parameter $\theta$ is given by the formula:

$$\theta \in [2\hat{\theta} - \hat{G}^{-1}(1-a),\ 2\hat{\theta} - \hat{G}^{-1}(a)], \tag{2.10}$$

where $G_B^{-1}(t)$ is the same as in (2.2).

## 3. CLASSICAL CONFIDENCE INTERVAL

The classical confidence interval (Fisher's method) for the correlation coefficient is extracted through the Z-transform, $\phi = 0.5\ln(\frac{1+\rho}{1-\rho}) = \tanh^{-1}(\rho)$.

The confidence interval for z is $\phi \in [\hat{\phi} - Z_\alpha \sigma(\phi),\ \hat{\phi} + Z_\alpha \sigma(\phi)] = [L, U]$

where $\hat{\phi} = \tanh^{-1}(r)$ and $\sigma^2(\phi) = 1/(n-3)$.

It follows that the confidence interval for the correlation coefficient is given by

$$\rho \in [(e^{2L} - 1)/(e^{2L} + 1),\ (e^{2U} - 1)/(e^{2U} + 1)] \tag{3.1}$$

The above formula was applied to both coefficients, since normality was assumed.

## 4. SOME CHARACTERISTICS OF THE ABOVE CONFIDENCE INTERVALS

The transformation respecting property allows us to construct confidence intervals for a parameter, then transform the endpoints of the interval and end up with confidence intervals for a transformation of the original parameter of interest. If, for instance, we have constructed a confidence interval for a parameter $\theta$, then the interval for $\sqrt{\theta}$ is derived straightforwardly by squaring the endpoints of the initial confidence interval. The accuracy term refers to the rate of convergence, of the coverage probability, to the desired level of coverage. A central (1-2α) confidence interval is supposed to have probability α of not covering the true value of $\theta$ from above and below. In the case of a sample this probability is equal to $α+\frac{c}{n}$ or $α+\frac{c}{\sqrt{n}}$ for some constant $c$. In the first case the fraction goes to zero at rate $1/n$, whereas in the second case it goes to zero at rate $1/\sqrt{n}$. We refer to the first case with the term second-order accuracy and with the term first-order accuracy to the second case.

The Normal (or Standard) method is known to be neither transformation respecting nor second-order accurate. Percentile and Basic methods are transformation respecting, but not second-order accurate. The BCα method is both transformation respecting and second-order accurate. The Studentized method is second-order accurate but not transformation respecting. If interest lies in estimating the correlation coefficient of the bivariate normal distribution then Fisher's transform works quite well. The problem arises in non normal populations like in our case. The drawback of not using a transformation is that one can end up with an interval not satisfying the range restriction. We did not use any transformation in the Studentized method and the result was obvious for small sized samples (n=10). The interval was larger than the set of permissible values for the correlation coefficient since this method is not range-preserving. The ABC method is both second-order accurate and transformation respecting. One could also say that since the Studentized method performs double bootstraps when the denominator is not known (in our case it is known), it needs more computational effort. On the other hand the ABC method requires far fewer replications than its counterpart (BCα) needs. More information on the advantages and disadvantages of the various bootstrap methods can be found in Hall (1988), Efron and Tibshirani (1993), and DiCiccio and Efron (1996).

## 5. SIMULATIONS AND RESULTS

A simulation study was performed as follows: 2000 independent samples of size 10, 20, 50 and 100 were generated from bivariate Poisson and bivariate Negative Binomial distributions. Each time we estimated the correlation coefficient using the two estimators and for the bootstrap confidence intervals construction, the number of bootstrap samples was set equal to B=1000. Bivariate Poisson variates were generated according to the method described in Morgan (1984). The probability mass function of the bivariate Poisson is given by the following formula:

$$P(X = x, Y = y) = \sum_{\delta=0}^{\min(x,y)} \frac{\lambda_1^{x-\delta} \lambda_2^{y-\delta} \lambda_3^{\delta}}{(x-\delta)!(y-\delta)!\delta!} \exp[-(\lambda_1 + \lambda_2 + \lambda_3)] \quad (5.1)$$

Parameters $\lambda_1$ and $\lambda_2$ were fixed at 0.5 and 1, respectively. What did not remain constant was the parameter of the covariance term ($\lambda_3$). The correlation between X and Y is given by the following formula

$$\rho = \frac{\lambda_3}{\sqrt{(\lambda_1 + \lambda_3)(\lambda_2 + \lambda_3)}} \quad (5.2)$$

For the selected values of the correlation, 0.25, 0.5, 0.75 and 0.9, the values of the covariance were 0.24, 0.73, 2.22 and 6.71, respectively.

Random values from a Negative Binomial distribution can be generated in many ways, by the process of compounding a Poisson distribution as in the univariate case or with the help of two univariate Negative Binomial variates and one univariate Binomial variate. These methods are better described in Kocherlakota and Kocherlakota (1992). In this paper the method of rejection sampling was used. For more information on the computer generation of the bivariate Negative Binomial distribution one can also look at Loukas and Kemp (1986). The distribution function is given by the formula

$$P(X = x, Y = y) = \frac{(r + x + y - 1)!}{(r-1)!x!y!} \cdot p_1^x p_2^y (1 - p_1 - p_2)^r \quad (5.3)$$

The correlation between X and Y is given by the following formula

$$\rho = \frac{\sqrt{p_1 p_2}}{\sqrt{(1-p_1)(1-p_2)}} \tag{5.4}$$

The values of the correlation coefficients were set equal to 0.25, 0.5 and 0.75. The confidence intervals for the correlation coefficient valued 0.9 in the bivariate Negative Binomial distribution were not computed due to computational difficulties. We set the parameters $p_1$ and $p_2$ to the values of (0.1393, 0.2786), (0.2287, 0.4574) and (0.2898, 0.5796), respectively. The number of successes ($r$) was set equal to 5. Bootstrap confidence intervals with coverage probability 1-2α=$\gamma_n$=0.95 for the correlation coefficient were calculated with each of the eight methods described above. Tables 1 & 3 and 2 & 4 present the expected coverage and average length of the confidence intervals for the Poisson and Negative Binomial cases, respectively.

### 5.1 Pearson's estimator

Tables 1 and 2 summarize the results of the simulations for Pearson's estimator. In the Poisson distribution, when the values of the correlation coefficient are less than or equal to 0.5 and the sample size is equal to 10, the average length exceeds unity. The Studentized method provides good confidence intervals in general (amongst the ones compared) but with the cost that the average length is the largest in all cases. The average length exceeded the length of the interval of possible values of the correlation coefficient. Since the pivotal quantity used here was not the same as Fisher's, this problem occurs naturally. However, as the sample size increases the estimated coverage probability reaches the nominal level of 0.95 faster and better than in the other methods. The Normal and Basic methods did not work very well in general. As correlation and sample size increase, they tend to provide better results.

The average lengths are in accordance with the lengths of the other methods, but this is not true for the estimated coverage. BCα and Fisher's based intervals are the most stable in general. No matter the value of the correlation and the sample size, they tend to produce stable results in terms of coverage probabilities. The ABC and Percentile methods perform well for large sample sizes but in general are not to be preferred as they tend to underestimate the true coverage probability, but not more than Normal and Basic methods. The Percentile method leads to similar conclusions.

**Table 1**. Estimates of the actual coverage, $\gamma_n = 0.95$ in percent (first line) and the expected length, (second line): Poisson distribution using Pearson's formula

| Correlations | Sample sizes | Methods | | | | | | | |
|---|---|---|---|---|---|---|---|---|---|
| | | Normal | Basic | Percentile | ABC | $BC\alpha(I.)$ | $BC\alpha(I.)$ | Studentized | Fisher's |
| $\rho=0.25$ | n=10 | 85.4 | 76.6 | 89.6 | 90.45 | 92.8 | 93.6 | 98.4 | 94.2 |
| | | 1.154 | 1.136 | 1.136 | 1.128 | 1.188 | 1.194 | -* | 1.126 |
| | n=20 | 87.35 | 82.7 | 91.85 | 91.2 | 92.95 | 93.5 | 95.8 | 93.15 |
| | | 0.826 | 0.819 | 0.819 | 0.794 | 0.826 | 0.829 | 1.064 | 0.829 |
| | n=50 | 90.95 | 88.95 | 93.0 | 92.4 | 92.5 | 92.85 | 94.15 | 92.9 |
| | | 0.534 | 0.533 | 0.533 | 0.526 | 0.534 | 0.534 | 0.584 | 0.534 |
| | n=100 | 91.8 | 90.4 | 92.85 | 93.15 | 92.65 | 93.05 | 93.6 | 92.9 |
| | | 0.386 | 0.386 | 0.386 | 0.383 | 0.387 | 0.386 | 0.404 | 0.386 |
| $\rho=0.5$ | n=10 | 84.35 | 75.4 | 90.1 | 87.74 | 92.35 | 93.8 | 96.55 | 92.75 |
| | | 1.007 | 0.987 | 0.987 | 0.987 | 1.053 | 1.077 | - | 1.077 |
| | n=20 | 88.55 | 82.5 | 92.3 | 91.4 | 93.45 | 94.2 | 95.85 | 93.4 |
| | | 0.683 | 0.677 | 0.677 | 0.666 | 0.6927 | 0.701 | 0.875 | 0.701 |
| | n=50 | 91.7 | 89.2 | 92.65 | 92.65 | 93.1 | 93.4 | 94.55 | 92.45 |
| | | 0.44 | 0.44 | 0.44 | 0.439 | 0.444 | 0.446 | 0.482 | 0.446 |
| | n=100 | 92.5 | 91.1 | 93.6 | 93.6 | 93.45 | 93.45 | 94.45 | 92.75 |
| | | 0.316 | 0.316 | 0.316 | 0.316 | 0.317 | 0.317 | 0.33 | 0.317 |
| $\rho=0.75$ | n=10 | 86.85 | 76.0 | 90.8 | 87.995 | 93.4 | 94.15 | 96.45 | 93.35 |
| | | 0.692 | 0.672 | 0.672 | 0.696 | 0.747 | 0.798 | - | 0.798 |
| | n=20 | 87.75 | 81.45 | 90.95 | 90.8 | 92.05 | 92.7 | 93.7 | 92.75 |
| | | 0.424 | 0.419 | 0.419 | 0.427 | 0.4407 | 0.455 | 0.54 | 0.455 |
| | n=50 | 91.55 | 88.15 | 93.55 | 93.45 | 93.45 | 93.75 | 95.0 | 92.85 |
| | | 0.263 | 0.263 | 0.263 | 0.267 | 0.269 | 0.272 | 0.285 | 0.272 |
| | n=100 | 92.5 | 89.95 | 93.65 | 92.9 | 93.0 | 93.25 | 93.55 | 92.15 |
| | | 0.187 | 0.187 | 0.187 | 0.19 | 0.19 | 0.191 | 0.195 | 0.191 |
| $\rho=0.9$ | n=10 | 87.7 | 73.3 | 89.4 | 84.2 | 91.6 | 92.5 | 95.25 | 93.15 |
| | | 0.372 | 0.35 | 0.35 | 0.377 | 0.419 | 0.473 | - | 0.473 |
| | n=20 | 88.45 | 81.7 | 91.75 | 90.2 | 91.65 | 92.1 | 93.5 | 93.8 |
| | | 0.2 | 0.197 | 0.197 | 0.208 | 0.214 | 0.225 | 0.25 | 0.225 |
| | n=50 | 91.5 | 86.05 | 93.7 | 93.65 | 93.5 | 93.65 | 94.4 | 93.4 |
| | | 0.116 | 0.116 | 0.116 | 0.12 | 0.12 | 0.122 | 0.125 | 0.122 |
| | n=100 | 92.6 | 90.7 | 93.55 | 93.85 | 94.4 | 94.35 | 94.15 | 92.9 |
| | | 0.082 | 0.082 | 0.082 | 0.084 | 0.084 | 0.085 | 0.085 | 0.085 |

* Length greater than 2.

**Table 2**. Estimates of the actual coverage, $\gamma_n = 0.95$ in percent (first line) and the expected length, (second line): Negative Binomial distribution using Pearson's formula

| Correlations | Sample sizes | Methods | | | | | | | |
|---|---|---|---|---|---|---|---|---|---|
| | | Normal | Basic | Percentile | ABC | BCα($I$) | BCα($I_+$) | Studentized | Fisher's |
| ρ=0.25 | n=10 | 86.0 | 78.0 | 92.0 | 88.45 | 93.4 | 94.4 | 98.6 | 94.6 |
| | | 1.19 | 1.173 | 1.173 | 1.151 | 1.209 | 1.218 | -* | 1.218 |
| | n=20 | 86.8 | 82.7 | 90.9 | 89.7 | 91.8 | 92.55 | 94.05 | 92.9 |
| | | 0.824 | 0.817 | 0.817 | 0.787 | 0.825 | 0.826 | 1.05 | 0.826 |
| | n=50 | 90.35 | 88.15 | 92.55 | 91.45 | 91.9 | 92.3 | 93.4 | 92.5 |
| | | 0.535 | 0.533 | 0.533 | 0.525 | 0.535 | 0.534 | 0.583 | 0.534 |
| | n=100 | 90.3 | 89.35 | 91.2 | 90.9 | 90.9 | 91.0 | 91.1 | 91.5 |
| | | 0.387 | 0.386 | 0.386 | 0.384 | 0.387 | 0.387 | 0.404 | 0.387 |
| ρ=0.5 | n=10 | 87.2 | 77.9 | 91.4 | 88.05 | 93.45 | 94.35 | 98.05 | 93.75 |
| | | 1.032 | 1.012 | 1.012 | 1.004 | 1.077 | 1.098 | - | 1.098 |
| | n=20 | 88.9 | 83.9 | 91.7 | 89.85 | 92.0 | 92.85 | 94.65 | 93.5 |
| | | 0.699 | 0.694 | 0.694 | 0.674 | 0.705 | 0.713 | 0.883 | 0.713 |
| | n=50 | 92.55 | 90.25 | 94.4 | 93.25 | 93.25 | 93.8 | 95.05 | 93.7 |
| | | 0.444 | 0.443 | 0.443 | 0.438 | 0.444 | 0.446 | 0.483 | 0.446 |
| | n=100 | 93.3 | 91.8 | 93.05 | 92.7 | 92.45 | 92.7 | 93.7 | 92.55 |
| | | 0.32 | 0.32 | 0.32 | 0.318 | 0.321 | 0.321 | 0.334 | 0.321 |
| ρ=0.75 | n=10 | 87.95 | 76.0 | 91.45 | 86.2 | 92.55 | 93.85 | 96.75 | 92.9 |
| | | 0.707 | 0.686 | 0.686 | 0.699 | 0.769 | 0.811 | *1.497* | 0.811 |
| | n=20 | 88.9 | 82.15 | 91.95 | 90.0 | 92.25 | 92.55 | 94.2 | 92.35 |
| | | 0.435 | 0.43 | 0.43 | 0.427 | 0.447 | 0.46 | 0.542 | 0.46 |
| | n=50 | 93.35 | 90.35 | 94.9 | 93.8 | 93.85 | 94.5 | 94.75 | 94.4 |
| | | 0.264 | 0.263 | 0.263 | 0.264 | 0.267 | 0.269 | 0.284 | 0.269 |
| | n=100 | 93.7 | 92.1 | 94.8 | 93.45 | 93.45 | 93.55 | 94.4 | 93.5 |
| | | 0.187 | 0.187 | 0.187 | 0.187 | 0.188 | 0.188 | 0.194 | 0.189 |

* Length greater than 2.

Similar conclusions are to be drawn in the bivariate Negative Binomial distribution case as well. The Normal and Basic methods perform the same as before and as the sample size increases from n=20 to n=50 the estimation of the coverage probability is closer to the desired nominal level but the average length is not reduced by the same amount as before. The Percentile method estimates the coverage to be more than 0.90 irrespective of the sample size or the correlation value. The performance of the ABC method is about the same as before. The stability of BCα and Fisher's methods are also met in this case. Studentized confidence intervals are more conservative; they overestimate the nominal level of 0.95 for small samples but later approximate the desired level.

Although these conclusions are drawn with respect to Pearson's formula, similar conclusions can be drawn when using Spearman's formula.

**5.2 Spearman's estimator**

Tables 3 and 4 summarize the simulation results for Spearman's estimator. In the bivariate Poisson case, when the values of the true correlation coefficient are less than or equal to 0.5 and the sample is of size 10, the average length of the 95% confidence intervals exceeds unity except for the ABC method with a correlation equal to 0.5. The estimated coverage tends to the nominal level as the values of the correlation and the sample size increase, but the convergence seems to be faster as the true values of the coefficient increase rather than as the sample size increases.

The Normal and Basic methods perform better with this non-parametric estimator and the Percentile method works much better in comparison to the parametric estimator regardless of the combinations of the sample size and the correlation coefficient values. The ABC method shows a significant improvement also. The BCα methods and Classical method (Fisher's transform) work very well under any circumstances, which was also the case before. However, BCα methods perform a little better using Spearman's estimator.

In contrast to Pearson's estimator, the Studentized method as applied in Spearman's estimator does not perform similarly. The coverage is approached for large samples only (from 50 and above) and as the correlation increases the approach is better.

With Negative Binomial distribution, things are slightly different. The average length exceeds unity in the same occasions as with the Poisson distribution. Normal and Basic methods using this non-parametric estimator perform slightly better than using the parametric estimator. The performance of the Percentile method is roughly at the same levels for both estimators, and so is the ABC method. The results for BCα, Studentized and Fisher's methods are similar using either estimator.

**Table 3**. Estimates of the actual coverage, $\gamma_n = 0.95$ in percent (first line) and the expected length, (second line): Poisson distribution using Spearman's formula

| Correlations | Sample sizes | Methods | | | | | | | |
|---|---|---|---|---|---|---|---|---|---|
| | | Normal | Basic | Percentile | ABC | $BC\alpha(I_{-})$ | $BC\alpha(I_{+})$ | Studentized | Fisher's |
| ρ=0.25 | n=10 | 88.8 | 80.6 | 93.2 | 89.92 | 95 | 95.2 | 80.6 | 94.2 |
| | | 1.2039 | 1.1805 | 1.1805 | 1.1375 | 1.2239 | 1.2337 | 1.1805 | 1.2367 |
| | n=20 | 90.45 | 86.9 | 93.05 | 90.7 | 94.65 | 94.8 | 86.9 | 92.75 |
| | | 0.8442 | 0.8396 | 0.8396 | 0.7951 | 0.8541 | 0.8553 | 0.8396 | 0.8553 |
| | n=50 | 93.4 | 91.85 | 94.5 | 92.35 | 95.5 | 95.55 | 91.85 | 92.35 |
| | | 0.5337 | 0.534 | 0.534 | 0.5263 | 0.5368 | 0.5369 | 0.534 | 0.5369 |
| | n=100 | 94.05 | 93.9 | 94.95 | 94.0 | 95.3 | 95.3 | 93.9 | 93.4 |
| | | 0.3777 | 0.3786 | 0.3786 | 0.3845 | 0.3794 | 0.3794 | 0.3786 | 0.3794 |
| ρ=0.5 | n=10 | 86.0 | 79.4 | 91.6 | 88.16 | 92.4 | 93.0 | 79.4 | 92.6 |
| | | 1.0551 | 1.0395 | 1.0395 | 0.9861 | 1.0994 | 1.1196 | 1.0395 | 1.1196 |
| | n=20 | 91.5 | 87.3 | 94.2 | 91.25 | 95.3 | 95.55 | 87.3 | 93.45 |
| | | 0.7274 | 0.7222 | 0.7222 | 0.6735 | 0.7507 | 0.7537 | 0.7222 | 0.7537 |
| | n=50 | 94.05 | 92.9 | 95.45 | 93.3 | 95.7 | 95.75 | 92.9 | 93.15 |
| | | 0.4525 | 0.4519 | 0.4519 | 0.4381 | 0.4597 | 0.4599 | 0.4519 | 0.4599 |
| | n=100 | 94.15 | 94.1 | 94.6 | 92.85 | 94.7 | 94.7 | 94.1 | 92.1 |
| | | 0.3178 | 0.3181 | 0.3181 | 0.3181 | 0.3209 | 0.321 | 0.3181 | 0.321 |
| ρ=0.75 | n=10 | 88.4 | 79.8 | 92.6 | 86.61 | 93.0 | 94.2 | 79.8 | 92.2 |
| | | 0.7848 | 0.7686 | 0.7686 | 0.6805 | 0.8456 | 0.8847 | 0.7686 | 0.8847 |
| | n=20 | 91.15 | 85.25 | 94.75 | 90.15 | 93.9 | 94.1 | 85.25 | 92.9 |
| | | 0.4969 | 0.4919 | 0.4919 | 0.4296 | 0.5246 | 0.5307 | 0.4919 | 0.5307 |
| | n=50 | 93.45 | 91.3 | 94.3 | 92.2 | 93.95 | 94.0 | 91.3 | 93.0 |
| | | 0.2965 | 0.2955 | 0.2955 | 0.267 | 0.3063 | 0.3068 | 0.2955 | 0.3068 |
| | n=100 | 94.2 | 93.6 | 93.4 | 93.3 | 93.2 | 93.15 | 93.6 | 92.9 |
| | | 0.2058 | 0.2058 | 0.2058 | 0.1879 | 0.2097 | 0.2098 | 0.2058 | 0.2098 |
| ρ=0.9 | n=10 | 92.6 | 80.0 | 96.8 | 86.48 | 93.8 | 95.0 | 80.0 | 94.0 |
| | | 0.4985 | 0.4824 | 0.4824 | 0.3846 | 0.5548 | 0.6076 | 0.4824 | 0.6076 |
| | n=20 | 94.0 | 86.9 | 96.35 | 90.15 | 93.8 | 94.2 | 86.9 | 93.45 |
| | | 0.2924 | 0.2871 | 0.2871 | 0.209 | 0.3097 | 0.3172 | 0.2871 | 0.3172 |
| | n=50 | 94.3 | 91.1 | 94.85 | 93.0 | 92.85 | 92.9 | 91.1 | 93.25 |
| | | 0.1552 | 0.1541 | 0.1541 | 0.122 | 0.1601 | 0.1608 | 0.1541 | 0.1608 |
| | n=100 | 95.6 | 94.3 | 94.45 | 93.2 | 92.9 | 92.85 | 94.3 | 92.95 |
| | | 0.1033 | 0.1031 | 0.1031 | 0.0843 | 0.1055 | 0.1056 | 0.1031 | 0.1056 |

**Table 4**. Estimates of the actual coverage, $\gamma_n = 0.95$ in percent (first line) and the expected length, (second line): Negative Binomial distribution using Spearman's formula

| Correlations | Sample sizes | Methods | | | | | | | |
|---|---|---|---|---|---|---|---|---|---|
| | | Normal | Basic | Percentile | ABC | BCα($I_-$) | BCα($I_+$) | Studentized | Fisher's |
| ρ=0.25 | n=10 | 84.6 | 78.2 | 90.0 | 85.57 | 92.6 | 93.4 | 78.2 | 93.0 |
| | | 1.2219 | 1.2034 | 1.2034 | 1.0786 | 1.2419 | 1.253 | 1.2034 | 1.253 |
| | n=20 | 90.8 | 87.8 | 94.15 | 90.6 | 95.35 | 95.45 | 98.15 | 93.3 |
| | | 0.8656 | 0.8601 | 0.8601 | 0.7859 | 0.8909 | 0.872 | 1.1189 | 0.872 |
| | n=50 | 92.85 | 92.4 | 93.95 | 91.8 | 94.4 | 94.4 | 92.4 | 93.75 |
| | | 0.5405 | 0.5402 | 0.5402 | 0.5265 | 0.5421 | 0.5422 | 0.5402 | 0.5422 |
| | n=100 | 91.35 | 91.35 | 91.6 | 91.35 | 91.8 | 91.8 | 91.35 | 91.65 |
| | | 0.3813 | 0.3822 | 0.3822 | 0.3822 | 0.3829 | 0.3829 | 0.3822 | 0.3829 |
| ρ=0.5 | n=10 | 89.7 | 82.75 | 95.6 | 88.38 | 95.8 | 96.6 | 82.75 | 93.55 |
| | | 1.1347 | 1.1169 | 1.1687 | 0.9877 | 1.1591 | 1.1802 | 1.1169 | 1.1802 |
| | n=20 | 92.45 | 89.55 | 95.05 | 89.85 | 95.1 | 95.2 | 98.15 | 93.3 |
| | | 0.7643 | 0.7593 | 0.7593 | 0.6816 | 0.7751 | 0.778 | 0.999 | 0.778 |
| | n=50 | 94.25 | 94.0 | 94.45 | 92.15 | 94.3 | 94.3 | 94.0 | 93.05 |
| | | 0.4671 | 0.4666 | 0.4666 | 0.4426 | 0.471 | 0.4713 | 0.4666 | 0.4713 |
| | n=100 | 93.8 | 94.3 | 92.55 | 92.8 | 92.05 | 92.05 | 94.3 | 92.45 |
| | | 0.3248 | 0.3251 | 0.3251 | 0.3251 | 0.3266 | 0.3267 | 0.3251 | 0.3267 |
| ρ=0.75 | n=10 | 93.85 | 85.7 | 96.8 | 86.31 | 96.15 | 96.9 | 85.7 | 93.6 |
| | | 0.8928 | 0.8743 | 0.8743 | 0.7222 | 0.9101 | 0.9482 | 0.8743 | 0.9482 |
| | n=20 | 95.25 | 90.35 | 96.3 | 90.5 | 95.3 | 95.55 | 90.35 | 93.2 |
| | | 0.5437 | 0.5388 | 0.5388 | 0.4364 | 0.5502 | 0.5567 | 0.5388 | 0.5567 |
| | n=50 | 95.6 | 93.9 | 94.5 | 91.7 | 94.1 | 94.05 | 93.9 | 92.7 |
| | | 0.3077 | 0.3071 | 0.3071 | 0.2654 | 0.3101 | 0.3107 | 0.3071 | 0.3107 |
| | n=100 | 95.45 | 94.9 | 93.35 | 93.4 | 92.95 | 92.95 | 94.9 | 93.1 |
| | | 0.2086 | 0.2084 | 0.2084 | 0.1875 | 0.2093 | 0.2095 | 0.2084 | 0.2095 |

**5.3 Examination of some properties of these two estimators under the given distributions**

The bias of the estimators is under examination in this section. We generated 1,000,000 pairs of data from each distribution using the 4 chosen correlation values (0.25, 0.5, 0.75 and 0.9) and estimated the correlation using both estimators. We repeated this procedure 1000 times. Finally, the mean of the 1000 estimators was compared to the true correlation value and the bias was extracted. The variance of these 1000 estimators was also calculated. The results are shown in Table 5.

As seen from Table 5 in the Poisson case, Pearson's estimator is asymptotically unbiased whereas Spearman's estimator is not. In fact, it always has a small negative bias. One element these two estimators have in common is their asymptotic normality. The distribution of both estimators for the sample examined was clearly non-normal (not even of symmetric form) and that was evident from their confidence intervals. Furthermore, Pearson's estimator converges to normality faster than his competitor's estimator. This was apparent since this procedure was repeated with 10,000 pairs of data and normality (using Shapiro's test) was rejected sometimes for Spearman's estimator, but never for Pearson's estimator.

**Table 5**. Estimated bias for large sample sizes (the variance is presented in parentheses)

| Correlations | Poisson distribution | | Negative Binomial distribution | |
| --- | --- | --- | --- | --- |
| | Pearson's estimation | Spearman's estimation | Pearson's estimation | Spearman's estimation |
| $\rho = 0.25$ | 0.2499 $(1.062 \times 10^{-5})$ | 0.2346 $(6.53 \times 10^{-6})$ | 0.2182 $(1.045 \times 10^{-4})$ | 0.1995 $(8.885 \times 10^{-5})$ |
| $\rho = 0.5$ | 0.5 $(6.9 \times 10^{-6})$ | 0.4817 $(6.49 \times 10^{-6})$ | 0.4823 $(6.67 \times 10^{-5})$ | 0.4568 $(6.23 \times 10^{-5})$ |
| $\rho = 0.75$ | 0.7499 $(2.29 \times 10^{-6})$ | 0.7354 $(2.48 \times 10^{-6})$ | 0.7458 $(2.31 \times 10^{-5})$ | 0.7267 $(2.41 \times 10^{-5})$ |
| $\rho = 0.9$ | 0.9 $(4.25 \times 10^{-7})$ | 0.8919 $(5.7 \times 10^{-7})$ | - | - |

**5.4 Comparison of the estimators in terms of the mean square error**

MSE is a criterion used to assess at some degree the efficiency of an estimator and to compare estimators. It is defined as the sum of the variance of the estimator and the squared bias of the estimator.

In this section, we compared the estimators using a wider range of values for the covariance parameter for each of the already studied sample sizes (n=10, n=20, n=50 and n=100). That is, we used values for the correlation coefficient ranging from 0.05 to 0.95, each time increasing by 0.01. The MSE for both estimators was estimated for all values using different sample sizes each time for both distributions. For every value of the correlation and each sample size, random values from each distribution were generated and the correlation coefficient was calculated 1000 times. Then the mean and the variance of these 1000 values were calculated and used for the extraction of the MSE. The results are shown in Figures 1 and 2 for the Poisson and negative Binomial, respectively.

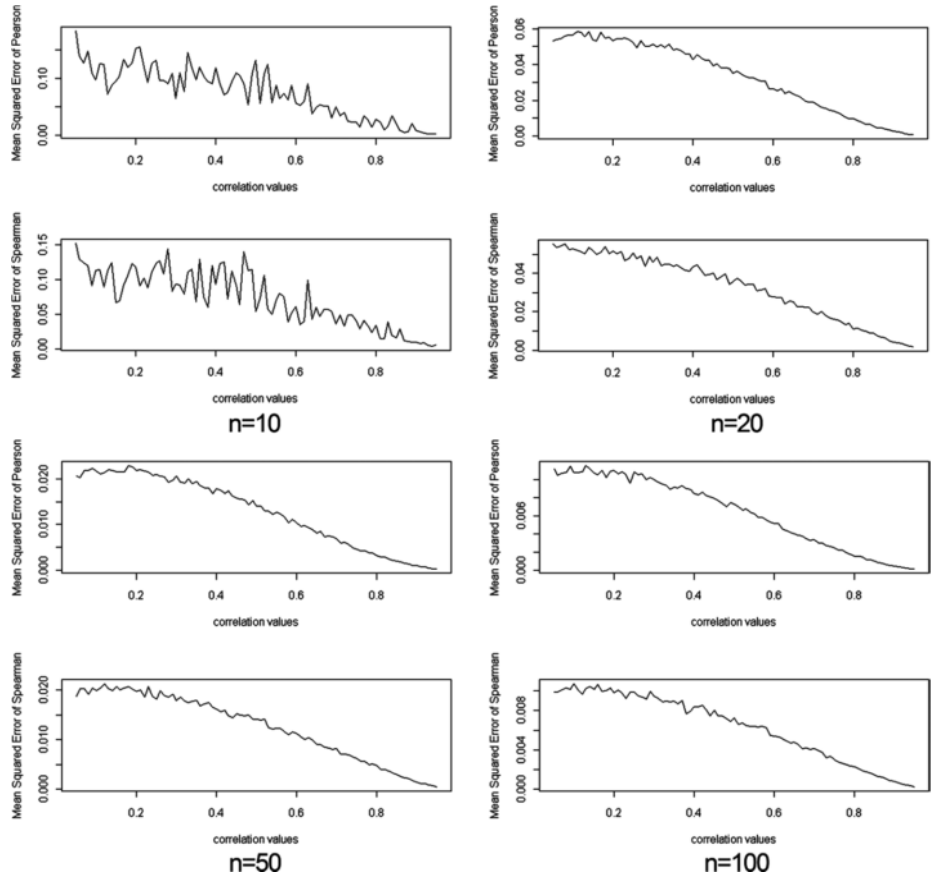

**Figure 1. Poisson distribution**

It can be seen that MSE values are very close to each other. Approximately half of the time, the difference between the MSE of Pearson's and the MSE of Spearman's estimator is positive. Regardless of the sign of the difference, the maximum difference between them does not exceed 0.003. In addition, for small values of the correlation, the MSE of Spearman's estimator is lower than that of Pearson's estimator, but as the correlation increases, the opposite pattern occurs.

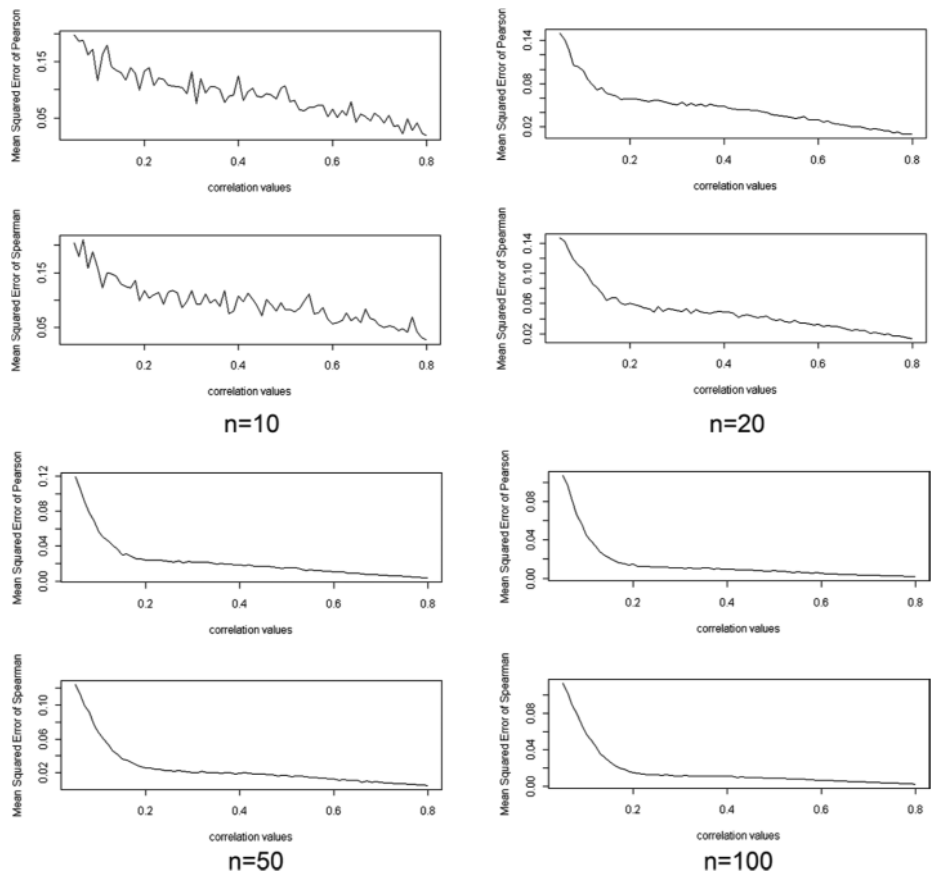

**Figure 2. Negative binomial distribution**

## 6. DISCUSSION

In this study an examination was performed of several bootstrap confidence intervals for the correlation coefficient when the populations are discrete. The two bivariate distributions that were examined were the Poisson and the Negative Binomial. Furthermore, two estimators were compared--Pearson's and Spearman's formula. What was apparent was that as the correlation between the two variables increases, the accuracy of the coverage probability also increases. The same is true with the sample size. As for the comparison of the confidence intervals, the BCα family of confidence intervals exhibited great stability under all circumstances. The same is true for Fisher's transformation, regardless of the estimator used (parametric or not).

MSE as a criterion for a further comparison of these two estimators showed that they produce results that are very close. A further examination though showed that Pearson's formula is asymptotically unbiased, whereas its non-parametric alternative is not. In addition, the parametric estimator tends to normality faster that the non-parametric estimator.

In our opinion, based upon the findings of the simulations we propose the use of Pearson's estimator instead of Spearman's and the Fisher's transform for confidence intervals construction. The reason is that Fisher's transformation is simpler than the Bcα, which also performs very well in general.

There are still more bootstrap techniques for the correlation coefficient and certainly many more bivariate distributions (discrete or continuous) whose correlation coefficients are to be examined.

**ACKNOWLEDGMENTS**

The authors would like to thank Constantinos C. Frangos, University College London, for his kind advice and support during the preparation of this manuscript. Additionally, we would like to thank Thodoris Kypraios, Lecturer at the University of Nottingham, for reading a first draft of this paper.